\documentclass[letterpaper, 10 pt, conference]{ieeeconf}  

\IEEEoverridecommandlockouts                              

\usepackage{amsfonts, dsfont, mathtools, amsmath, amssymb, bbold, mathrsfs, bm, upgreek}
\usepackage{graphicx, float, epsfig, color, psfrag, adjustbox, enumerate}
\usepackage[font=small]{caption}
\usepackage{refcount, url, cleveref}
\usepackage[noadjust]{cite}
\usepackage[usenames,dvipsnames,svgnames,table]{xcolor}

\title{\LARGE \bf
A Sensory Feedback Control Law for Octopus Arm Movements
}

\author{Tixian Wang$^{1,2}$, Udit Halder$^2$, Ekaterina Gribkova$^3$, \\ [5pt]
 Rhanor Gillette$^{3,4}$, Mattia Gazzola$^{1,5,6}$, Prashant G. Mehta$^{1,2}$
\thanks{We gratefully acknowledge financial support from ONR MURI N00014-19-1-2373, and NSF EFRI C3 SoRo $\#$1830881. We also acknowledge computing resources provided by the Extreme Science and Engineering Discovery Environment (XSEDE), which is supported by National Science Foundation grant number ACI-1548562, through allocation TG-MCB190004.}%
\thanks{$^{1}$Department of Mechanical Science and Engineering, $^{2}$Coordinated Science Laboratory, $^{3}$Neuroscience Program, $^{4}$Department of Molecular and Integrative Physiology,
$^{5}$National Center for Supercomputing Applications, $^{6}$Carl R. Woese Institute for Genomic Biology, University of Illinois at Urbana-Champaign. 
  Corresponding e-mail:  {\tt\small udit@illinois.edu}}%
}
\def\R{{\mathds{R}}}

\def\0{{\mathbb{0}}}
\def\1{{\mathds{1}}}

\DeclareMathOperator*{\argmin}{arg\,min}

\definecolor{db}{RGB}{23,20,119}
\definecolor{dg}{RGB}{2,101,15}

\newtheorem{proposition}{Proposition}[section]

\newtheorem{remark}{Remark}

\usepackage{upgreek}
\newcommand{\dif}{\mathrm{d}}

\newcommand{\set}[1]{\left\{#1\right\}}

\newcommand{\material}[1]{
	\ifthenelse{\equal{#1}{\kappa}}{\upkappa}{
	\ifthenelse{\equal{#1}{\nu}}{\upnu}{
	\ifthenelse{\equal{#1}{\omega}}{\upomega}{
	\ifthenelse{\equal{#1}{\sigma}}{\upsigma}{
	\ifthenelse{\equal{#1}{\theta}}{\uptheta}{
	\mathsf{#1}}}}}}
}

\newcommand{\target}{\text{target}}

\newcommand{\longitudinalmuscle}{\text{LM}}

\newcommand{\ud}{\,\mathrm{d}}

\newcommand{\myper}{_\text{per}}
\newcommand{\mytan}{_\text{tan}}

\newcommand{\rhow}{\rho_\text{water}}

\newcommand{\radius}{\gamma^\text{rod}}
\newcommand{\rodtip}{\gamma^\text{tip}}
\newcommand{\rodbase}{\gamma^\text{base}}

\newcommand{\dist}{\rho}
\newcommand{\distvec}{\boldsymbol\rho}

\newcommand{\unicycle}[1]{\mathsf{#1}}
\newcommand{\thetaUnicycle}{\vartheta}
\newcommand{\slope}{\mathsf{m}}

\begin{document}
\bstctlcite{BSTcontrol} 
\maketitle
\thispagestyle{empty}
\pagestyle{empty}


\begin{abstract}
The main contribution of this paper is a novel sensory feedback control law for an octopus arm.  The control law is inspired by, and helps integrate, several observations made by biologists.  
The proposed control law is distinct from prior work which has mainly focused on open-loop control strategies.  Several analytical results are described including characterization of the equilibrium and its stability analysis. Numerical simulations demonstrate
life-like motion of the soft octopus arm, qualitatively matching behavioral experiments. Quantitative comparison with bend propagation experiments helps provide the first explanation of such canonical motion using a sensory feedback control law.  
Several remarks are included that help draw parallels with natural pursuit strategies such as motion camouflage or classical pursuit.

\end{abstract}



\section{Introduction} \label{sec:intro}
Octopus arm movements have been studied extensively by the biologists during the past few decades \cite{gutfreund1998patterns, sumbre2001control, sumbre2005motor, kennedy2020octopus}.  
Octopus arms are hyper-flexible and have virtually infinite degrees of freedom.  Although this flexibility yields an impressive repetoire of motions, it also makes discerning the underlying sensorimotor control mechanisms a challenging task. Several hypotheses and methods have been proposed for the control of octopus arms, including stiffening wave actuation~\cite{yoram2002move,yekutieli2005dynamic,yekutieli2005dynamic2,wang2022control}, energy shaping control~\cite{chang2020energy,chang2021controlling}, and optimal control~\cite{cacace2019control, wang2021optimal}. These earlier model-based studies typically lack the integration of sensory information into the motor control. How octopuses use their sensing capabilities to control their arms remains an open question.

Sensorimotor control has been widely investigated in other animals. Biologically plausible control laws for pursuit have been proposed, such as motion camouflage (observed in bats, dragonflies, falcons)\cite{glendinning2004mathematics, ghose2006echolocating,justh2006steering,kane2014falcons} or classical pursuit (observed in honey bees, flies) \cite{pj2007chases, wei2009pursuit, galloway2013symmetry, halder2016steering}. The majority of these studies focus on terrestrial or aerial creatures. Little has been put forth on the mathematics of potential pursuit strategies of octopuses. 

The main contribution of this paper is a novel sensory feedback control law for octopus arms.  The control law is inspired by several behavioral observations and biophysical experiments.



\subsection{Biological inspiration for sensorimotor control}\label{sec:inspiration}

\begin{figure*}[t]
	\centering
	\includegraphics[width=\textwidth, trim = {0pt 0pt 0pt 0pt}]{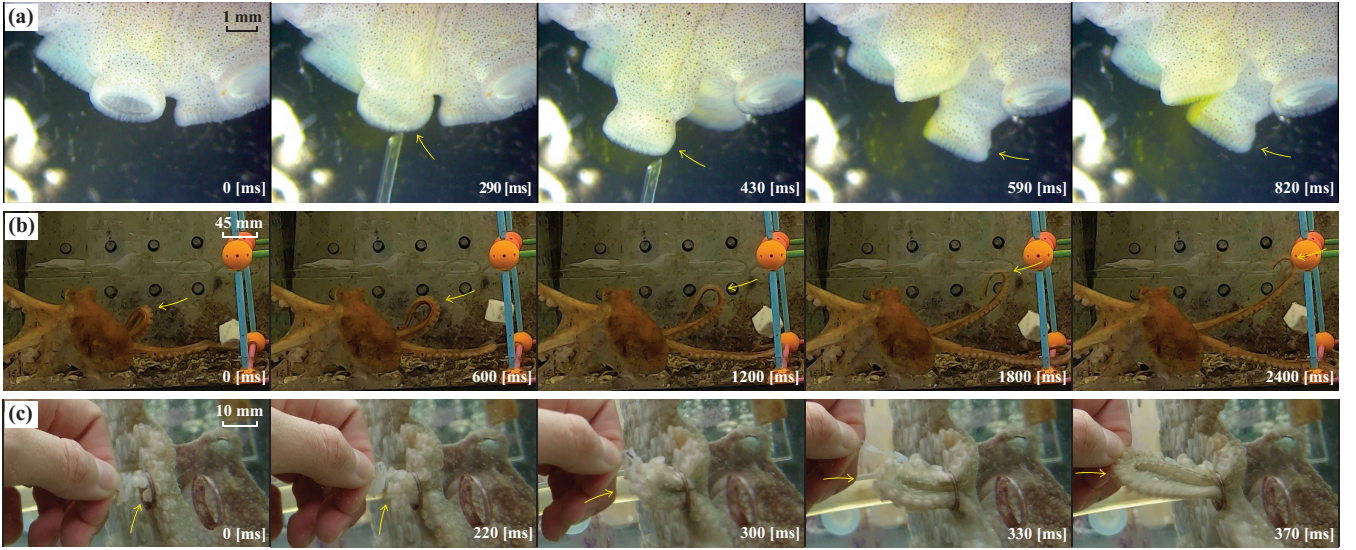}
	\caption{(a) A recording of suckers of an isolated octopus arm reacting to stimuli by extending and changing their orientation. The yellow arrows indicate the moving sucker. (b) A video sequence of an octopus performing bend propagation. The yellow arrows indicate the location of the bend. (c) A video recording of an octopus sensing a food source and reaching towards the target by squeezing through a hole. The yellow arrows indicate the location of the bend.}
	\label{fig:exp}
	\vspace{-15pt}
\end{figure*}

\noindent
{\bf 1) Local target bearing sensing through suckers:} An octopus arm is equipped with an array of suckers on one of its sides. Each sucker has an abundance of sensory receptors (tens of thousands), including chemosensory and mechanosensory cells \cite{graziadei1976sensory, mather2021octopus}, to detect a variety of stimuli.
Fig.~\ref{fig:exp}(a) depicts successive video frames from a laboratory experiment where shrimp extract is pipetted in the proximity of a group of suckers in an isolated octopus arm.
Frames show that suckers change their orientation in response to the chemical stimulus and reach towards the source. These experiments indicate that a sucker is able to estimate the bearing to a target stimulus.

\smallskip
\noindent
{\bf 2) Bend propagation:} Bend propagation in an octopus arm refers to a stereotypical maneuver whereby an octopus pushes a bend (localized region of large curvature) from the base to the tip of the arm (Fig.~\ref{fig:exp}(b)).  It is the most widely studied motion primitive in an octopus arm~\cite{gutfreund1998patterns, sumbre2001control, yekutieli2005dynamic, yoram2002move}.  Another related motion pattern is depicted in Fig.~\ref{fig:exp}(c), where a food source (shrimp) is presented to the octopus on the other side of a glass obstacle with holes. The octopus near the hole first senses the food source with suckers. Then it squeezes its arm through the hole by creating a bend which is subsequently propagated to catch the food.

\smallskip
\noindent
{\bf 3) The arm is passive beyond the bend point:} Electromyogram (EMG) recordings indicate that during a bend propagation maneuver, octopuses engage a wave of muscle actuation from the base to the tip \cite{gutfreund1998patterns}. In particular, it has been suggested that the arm muscles are activated from the base to the point where the bend is formed, while the remaining portion of the arm (bend to tip) stays passive. Our recent work \cite{wang2022control} on an open-loop control scheme that mimics this type of muscle control was found to reproduce the bend propagation patterns. 

\subsection{Contributions}
This paper builds on a body of work from our group \cite{chang2020energy, chang2021controlling, wang2022control} on control-oriented modeling of soft octopus arm and its internal musculature based on the Cosserat rod theory.  The unique contributions of this paper are as follows:

\smallskip
\noindent
1) A sensor model is proposed based on experimental observations involving suckers (Fig.~\ref{fig:exp}(a)).
Based on the sensor model, a biologically plausible feedback control law is proposed. The control is in the form of internal muscle couples whose magnitude depends on the target bearing and the location of the arm closest to the target. 

\smallskip
\noindent
2) An analysis of the equilibrium is described to show that stationary targets in the workspace are reached. A stability characterization of the equilibrium is also obtained.  Finally, motion patterns (such as bend propagation) that do not necessarily involve reaching a target are shown to be encompassed within the framework of our proposed control scheme.

	
\smallskip
\noindent
3) A numerical comparison is provided against the experimental data of bend propagation in a freely moving octopus arm.  The control law is shown to reproduce bell shaped bend velocity profiles.  The comparison helps provide the first such explanation of the bend propagation using a feedback control law.  

		
\smallskip
\noindent
4) We also report additional numerical simulation results which show life-like motion of the soft arm for a range of observed octopus behaviors including bend formation, bend propagation, and interaction with obstacles.  

\smallskip
The remainder of this paper is organized as follows. The dynamic model of an elastic rod is presented in Sec.~\ref{sec:model}. The sensory model and the feedback control law are introduced in Sec.~\ref{sec:sensing_ctrl}. An analytical study of the proposed control law is provided in Sec.~\ref{sec:analysis}. Control results are demonstrated by numerical simulations in Sec.~\ref{sec:numerics}, followed by a comparison with bend propagation experiments in Sec.~\ref{sec:experiment}.  Conclusions appear in Sec.~\ref{sec:conclusion}.

\section{Mathematical Model of an Octopus Arm} \label{sec:model}
A soft octopus arm is modeled as a planar Cosserat rod \cite{antman1995nonlinear, chang2020energy, chang2021controlling}. For simplicity of analysis, we consider the rod to be inextensible and unshearable (Kirchhoff rod).  Let $\set{{\mathbf{e}}_1,\mathbf{e}_2}$ denote a fixed orthonormal basis for the two-dimensional laboratory frame. The independent variables are the time $t\in\R$ and the arc-length $s\in[0,L_0]$ where $L_0$ is the length of the undeformed rod (see Fig.~\ref{fig:sensor}). The subscripts $(\cdot)_t$ and $(\cdot)_s$ denote the partial derivatives with respect to $t$ and $s$, respectively. 

The position vector of the centerline is denoted by $\mathbf{r}(s,t) \in \R^2$ and the angle $\theta(s,t) \in[0,2\pi)$ describes the material frame spanned by the orthonormal basis $\set{\mathbf{a}, \mathbf{b}}$, where $\mathbf{a} = \cos \theta \,\mathbf{e}_1 + \sin \theta \, \mathbf{e}_2, ~ \mathbf{b} = -\sin \theta \, \mathbf{e}_1 + \cos \theta \, \mathbf{e}_2$. The vector $\mathbf{a}$ is defined to be normal to the cross section. The kinematics of the rod are given by the following equations
\begin{equation}
	\mathbf{r}_s = \begin{pmatrix} \cos \theta \\ \sin \theta \end{pmatrix} = \mathbf{a}, \quad \theta_s = \kappa
	\label{eq:kinematics}
\end{equation}
The dynamics of a muscular octopus arm are described by a set of partial differential equations which require specification of internal passive elastic stresses, giving rise to the forces $\mathbf{n}$ and couples $m$, as well as internal active forces and couples generated by muscles. The internal forces $\mathbf{n} = n_1 \mathbf{a} + n_2 \mathbf{b}$ are to be determined so as to satisfy the rod's kinematic constraints of inextensibility and unshearability. We adopt the linear stress-strain relationship $m = EI \kappa = EI \theta_s$ for the internal couple, where $E$ is the Young's modulus of the arm and $I$ is the second moment of area of its cross section. We then write down the simplified dynamics of the muscular arm as follows:
\begin{equation}
	\begin{aligned}
		(\varrho A \mathbf{r}_t)_t &= (\mathsf{Q}\mathbf{n})_s - \zeta \mathbf{r}_t + \mathbf{f}^{\text{drag}} \\
		(\varrho I \theta_t)_t &= (EI \kappa)_s + n_2  - \zeta \theta_t + u_s 
	\end{aligned}
	\label{eq:dynamics}
\end{equation}
where $\mathsf{Q} = [\mathbf{a} ~~ \mathbf{b} ]$ 
is the planar rotation matrix, $\varrho$ is density, $A$ is the cross sectional area,  and $\zeta > 0$ is a damping coefficient which models viscoelastic dissipation within the arm. The effect of drag forces due to the surrounding fluid environment is modeled through the term $\mathbf{f}^{\text{drag}}$, which is explained in Appendix~\ref{appdx:drag}. Since the arm is assumed to be inextensible and unshearable,  without loss of generality the muscle actuations can be simplified to a couple control, denoted by $u$. The dynamics \eqref{eq:dynamics} are accompanied by a fixed-free boundary condition 
\begin{align}
\mathbf{r}(0, t) = 0,~ \theta(0, t) = 0, ~\mathbf{n}(L_0, t) = 0,~ m(L_0, t) = 0
\label{eq:boundary_conditions}
\end{align}

\begin{remark}
Since the muscle actuation is internal, we write the effective external couple as $u_s$ in \eqref{eq:dynamics}. For our numerical simulations, we adopt a biophysically realistic muscle model, as established in \cite{chang2021controlling}. In particular, we consider two longitudinal muscles (top and bottom, see Fig.~\ref{fig:sensor}). These muscles run along the length of the arm and are responsible for bending the arm. 
 Let $u^{\longitudinalmuscle_\mathsf{t}}$ and $u^{\longitudinalmuscle_\mathsf{b}}$ denote the muscle actuation for top and bottom longitudinal muscles, respectively. 
 Muscle actuations as a function of the couple control $u$ are then given by
\begin{equation*}
u^{\longitudinalmuscle_\mathsf{t}} = u  {\mathds{1}\{u\geq0\}}, \quad
u^{\longitudinalmuscle_\mathsf{b}} = u  {\mathds{1}\{u<0\}}
\end{equation*}
where $\mathds{1}\{\cdot\}$ is the indicator function.
\end{remark}

\section{Sensorimotor Control}\label{sec:sensing_ctrl}
Consider a target (food source) located at $\mathbf{r}^\target \in \R^2$. The arm senses the chemical signal emanated by the food source. The control problem is to devise a control strategy based on the sensory information to reach the target. 

We define
\begin{equation}
	\distvec(s,t) = \mathbf{r}^\target - \mathbf{r}(s,t),\quad \dist(s,t) = |\distvec(s,t)|
	\label{eq:def-dist}
\end{equation}
to be the vector and the Euclidean distance, respectively, from every point along the rod to the target. The bearing angle to the target with respect to the tangent vector along the rod is denoted as $\alpha$, so that
\begin{equation}
	\mathsf{R}(\alpha(s,t)) \frac{\mathbf{r}_s(s,t)}{|\mathbf{r}_s(s,t)|} = \frac{\distvec(s,t)}{\dist(s,t)}
	\label{eq:def-angle}
\end{equation}
where $\mathsf{R}(\alpha)$ is the planar rotation matrix for rotating a vector in $\R^2$ counterclockwise by the angle $\alpha$.

\subsection{Sensory model}\label{sec:sensor}

In this paper, we assume the octopus arm has, at all times, access to the following (chemo)sensory information:

\smallskip
\noindent
1) Bearing information at each location along the arm $\alpha(s,t)$.

\smallskip
\noindent
2) Arch-length of the closest point to the target, which we denote by $\bar{s}(t) \in [0, L_0]$ so that $\bar{s}(t):=\underset{s \in [0, L_0]}{\argmin}~ \dist(s,t)$. 
%

\begin{figure}[t]
	\centering
	\includegraphics[width=\columnwidth,trim={0pt 0pt 0pt 0pt}]{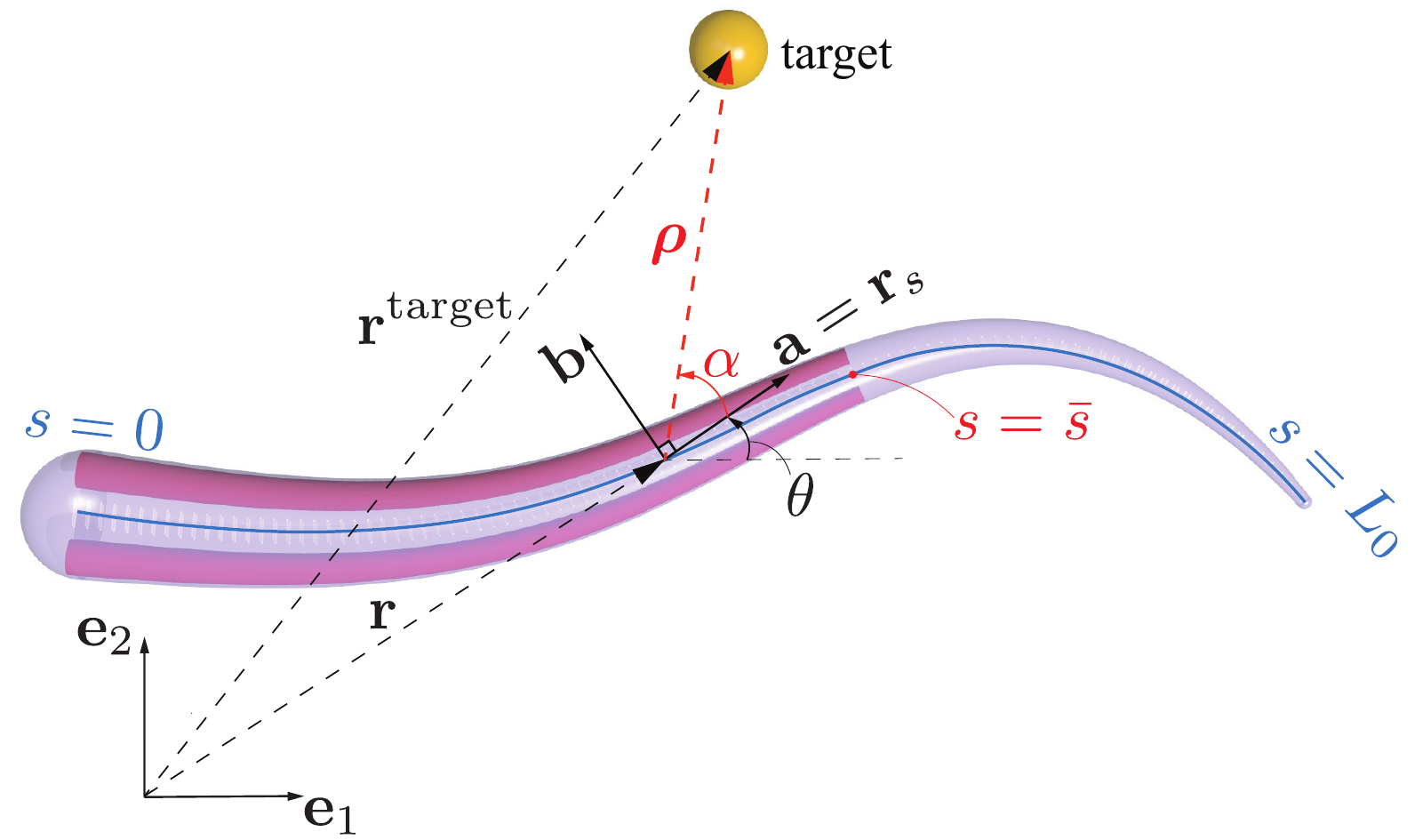} 
	\caption{Schematic of the sensorimotor control: The sensory information includes the angle $\alpha$ between the target vector $\distvec$ and the tangent vector $\mathbf{r}_s$, and the arch-length $\bar{s}$ of the closest point to the target. The control is in the form of internal muscle couple (illustrated in red) that actuates the arm from the base ($s=0$) to the closest point ($s=\bar{s}$).}
	\label{fig:sensor}
	\vspace{-10pt}
\end{figure}

\medskip
\begin{remark}
Estimating the bearing angle $\alpha$ from sensed chemical concentration is a non-trivial problem and is out of the scope of the current work. Future work will incorporate biologically informed (chemo)sensing models \cite{yafremava2011putative, gardner2000coding}. Moreover, the $\bar{s}$ point may be interpreted as the point of the arm receiving maximum chemical concentration. 

\end{remark}

\subsection{A biologically plausible feedback control law}
Based on the experimental observations described in Sec.~\ref{sec:inspiration} and the above sensory model, we propose the arm's feedback control law as 
\begin{equation}
	u(s,t) = -\mu(s,t) \sin(\alpha(s,t))
	\label{eq:ctrl_law_rod}
\end{equation}
where $\mu(s,t) \geq 0$ is a gain function, written as
\begin{equation}
	\mu(s,t) = \tilde{\mu} EI(s) \mathds{1}\{s \leq \bar{s}(t)\}
\label{eq:ctrl_law_mu}	
\end{equation}
where $\tilde{\mu} >0$ is a constant to be chosen. The indicator function in \eqref{eq:ctrl_law_mu} renders the rod passive beyond the $\bar{s}$ point (see Fig.~\ref{fig:sensor}). 


\begin{remark}
The inspiration behind the proposed control law comes from models of pursuit strategies such as classical pursuit or motion camouflage. In particular, one might find certain similarities between the feedback control law \eqref{eq:ctrl_law_rod} and the motion camouflage control law \cite[Sec.~3]{justh2006steering}. These connections are elucidated in the subsequent sections.  
\end{remark}

%

\section{Analysis} \label{sec:analysis}
In this section, we provide analytical results regarding the behavior of the arm under the proposed feedback control law. For simplicity, here we only consider a stationary target.

We express the spatial variation of the distance and angle to the target $(\dist, \alpha)$ by using the kinematics of the rod \eqref{eq:kinematics} and the definitions \eqref{eq:def-dist}-\eqref{eq:def-angle} as 
\begin{equation}
\begin{aligned}
\dist_s (s,t) &= -\cos(\alpha (s,t))  \\
\alpha_s (s,t) &= -\kappa(s,t) + \frac{1}{\dist (s,t)}\sin(\alpha (s,t)) 
\end{aligned}
\label{eq:sensory_kinematics}
\end{equation}

\begin{remark} 
\textit{Analogy with planar pursuit:} Note the similarity between differential equations~\eqref{eq:sensory_kinematics} in spatial domain and governing equations in time domain of a unicycle pursuing a stationary target (see the brief discussion in Appendix~\ref{appdx:pursuit} and compare with equations \eqref{eq:bearing_dynamics}). This gives an opportunity to draw a parallel, by interchanging the spatial variable ($s$ for rod) and the temporal variable ($t$ for pursuit trajectory). Consider a virtual agent, initialized at the origin and oriented toward the $\mathbf{e}_1$-axis (analogue to fixed boundary conditions at the base of the rod $\mathbf{r}(0) =0, \theta(0) = 0$), moving with a constant unit speed (analogue to constant unit stretch of the rod due to inextensibility and unshearability constraints). Assume steering control of the agent is given by the curvature $\kappa$ of the rod. Then the temporal trajectory of such a virtual agent would be exactly the same as the spatial configuration of the rod. 
\label{remark:analogy}
\end{remark}

In the light of this analogy, we investigate the equilibrium of the rod under the proposed feedback control law, and in particular its relation to the target. 

\subsection{Equilibrium analysis}
Any equilibrium of the arm must satisfy the equations of statics that are obtained from the dynamics \eqref{eq:dynamics} and the boundary conditions \eqref{eq:boundary_conditions} as
\begin{align*}
(EI(s) \kappa(s))_s + (u(s))_s = 0
\end{align*}
Under the proposed feedback control law, an equilibrium is found by solving the following equation for the curvature $\kappa$
\begin{align}
\kappa (s) = \frac{\mu (s)}{EI(s)}\sin(\alpha (s))
\label{eq:kappa_equilibrium}
\end{align}
Notice the similarity between \eqref{eq:kappa_equilibrium} and the steering control law for motion camouflage \eqref{eq:ctrl_law_MC}. This implies the equilibrium configuration of the rod can be viewed as a motion camouflage trajectory.   


At the equilibrium, plugging \eqref{eq:kappa_equilibrium} into \eqref{eq:sensory_kinematics} yields the closed-loop system
\begin{equation}
\begin{aligned}
\dist_s (s) &= -\cos(\alpha(s)) \\
\alpha_s (s) &= \left(\frac{1}{\dist(s)}-\frac{\mu(s)}{EI(s)}\right)\sin(\alpha(s))
\end{aligned}
\label{eq:sensory_kinematics_closed_loop}
\end{equation}

The equilibrium of the rod has two notable properties that are described by the following proposition.


%
%

\medskip
\begin{proposition} \label{prop:catching}
(i) Suppose the target location is such that $\dist(0) \leq L_0$.
Then for all $\epsilon >0$ there exists a $\tilde{\mu}>0$ large enough such that $\dist(\bar{s}) \leq \epsilon$ for some $\bar{s} \in [0, L_0]$. Physically this corresponds to the arm reaching the target.

(ii) Suppose the target location is such that $\dist(0) > L_0$.
Then for all $\epsilon > 0$ there exists a $\tilde{\mu}>0$ large enough such that $\cos (\alpha(L_0)) \geq 1 - \epsilon$. Physically this corresponds to the arm pointing toward the target.
\end{proposition}
\smallskip
\noindent
A proof of the proposition is provided in Appendix~\ref{appdx:catching}. Numerical simulations that demonstrate these two cases are given in Sec.~\ref{sec:numerics_static_target}.
\smallskip
\begin{remark}
	\textit{Time optimality of motion camouflage:} It has been shown that the motion camouflage law is a time-optimal strategy to capture a target \cite{ghose2006echolocating} ($t$ is minimized). According to the above analogy between the temporal variable in motion camouflage and the spatial variable of the rod configuration, we conclude that the location of the arm ($\bar{s}$) that reaches the target is minimized in $[0, L_0]$. This leads to a hypothesis that muscle energy expenditure is minimized under the proposed control law, which is meaningful biophysically. However, a rigorous analysis and supporting experimental study is a subject of our future work. 
\end{remark}

\subsection{Dynamic analysis}
\begin{proposition}
Consider the dynamics of the arm \eqref{eq:dynamics} with the feedback control law \eqref{eq:ctrl_law_rod}-\eqref{eq:ctrl_law_mu}. Then the equilibrium defined by \eqref{eq:kappa_equilibrium} is (locally) asymptotically stable.
\label{prop:equilibrium}
\end{proposition}
\smallskip
A proof of the proposition is given in Appendix~\ref{appdx:equilibrium}.

\smallskip
\begin{remark}
It is also of interest to analyze the trajectory of the closest point $\bar{s}(t)$ during the pursuit maneuver. Intuitively, the closest point moves directly toward the target, or in other words it follows a classical pursuit trajectory \cite{galloway2013symmetry}. We do not include analysis of such kind in this paper on account of space, but a numerical study is provided in Sec.~\ref{sec:simulation_moving_target}. 
\label{remark:dynamic_CP}
\end{remark}

\subsection{Special case: target at $\infty$}
It has been pointed out that the octopuses sometimes seem to be unable to correct the reaching movements after their initiation of a bend propagation pattern~\cite{yoram2002move}. This means that bend propagation may be encoded in the octopus sensorimotor control system as a primitive motion (not target-oriented). In this subsection, we show that this type of primitive motion can be obtained as a special case of our proposed feedback control law.

Assume that the target position is parameterized by the slope $\slope$ as $\mathbf{r}^\target=(x^*,y^*)=(x^*,\slope x^*)$. Denote the angle $\varphi$ to be the orientation of the target vector $\distvec$, so that
\begin{equation*}
	\tan \varphi=\frac{y^*-y}{x^*-x}=\frac{\slope-y/x^*}{1-x/x^*}
\end{equation*}
Now let the target go to infinity by taking $x^*\rightarrow\infty$. This yields $\varphi\rightarrow\tan^{-1}(\slope)$ for every point on the rod. It is also trivial to see that $\theta(\bar{s}) \rightarrow \tan^{-1}(-\frac{1}{\slope})$. We also have
\begin{equation}
	\begin{aligned}
		\sin\alpha = \sin(\varphi - \theta) = \frac{1}{\sqrt{1+\slope^2}}\left(\slope\cos\theta - \sin\theta \right)
	\end{aligned}
\label{eq:theta_feedback}
\end{equation}


In this case we can explicitly calculate the equilibrium configuration. The equilibrium according to~\eqref{eq:kappa_equilibrium} is now written as
\begin{equation*}
\theta_s = \frac{\mu}{EI}\frac{1}{\sqrt{1+\slope^2}}(\slope\cos\theta-\sin\theta)
\end{equation*}
whose exact solution is given by
\begin{equation*}
	\begin{aligned}
		\theta(s) &=  2\tan^{-1}\left(\frac{\sqrt{1+\slope^2}}{\slope}\tanh\left(\frac{1}{2}\int_0^s\tilde{\mu}\mathds{1}\{s'\leq \bar{s}\}\ud s'  \right. \right. \\ 
		&\qquad\qquad\quad \left. \left. + \tanh^{-1}\left(\frac{1}{\sqrt{1+\slope^2}} \right) \right) - \frac{1}{\slope} \right)
	\end{aligned}
\end{equation*}

\begin{remark}
Given the slope $\slope$, the control law~\eqref{eq:ctrl_law_rod}-\eqref{eq:ctrl_law_mu} does not depend on the chemosensory information but is instead a state feedback law on $\theta$ only. The slope $\slope$ serves as an estimate of the overall direction of the (virtual) target, which is chosen before instigating the bend propagation primitive. The feedback on $\theta$ indicates that local proprioceptive information (curvature) is needed for carrying out the bend propagation movement. Studies \cite{graziadei1965muscle, gutnick2020use} have found existence of proprioceptive cells around the intramuscular nerve cords of the arms which are consistent with such a hypothesis.
\end{remark}

\begin{figure*}[!t]
	\centering
	\includegraphics[width=\textwidth, trim = {0pt 0pt 0pt 0pt}]{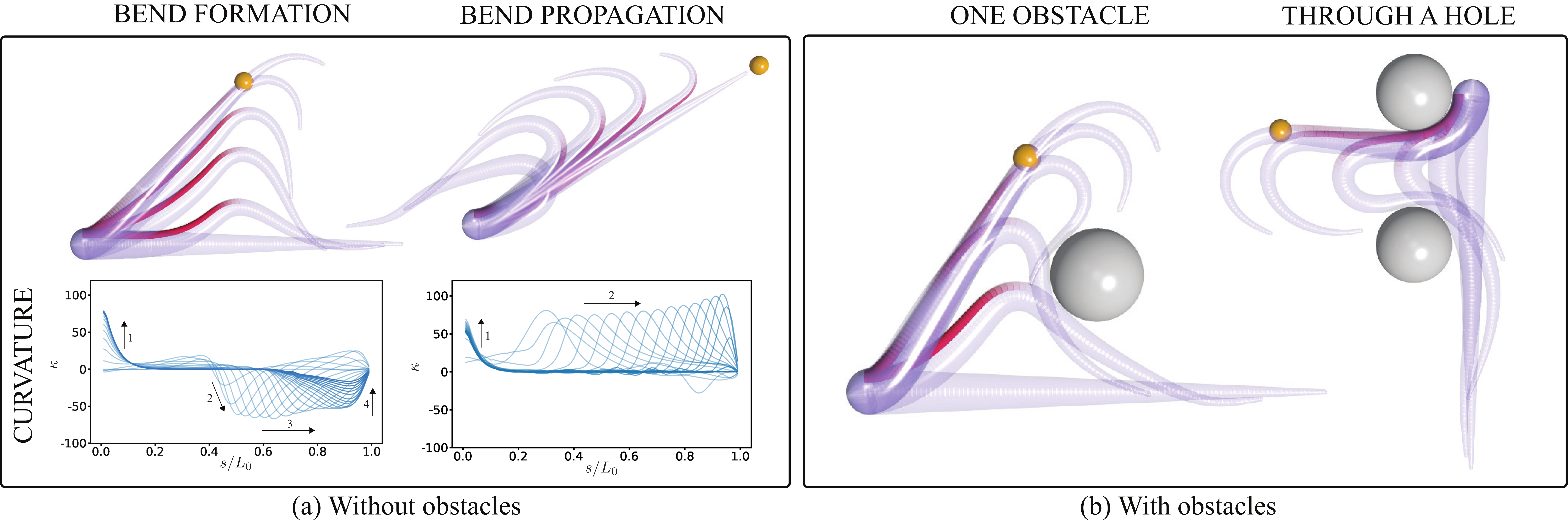}
	\caption{Simulation results of rod reaching a static target: We select six time instances for each rod configurations. The rod is shown in faded purple with the muscle actuations illustrated in red for longitudinal muscles. A sequence of time snapshots are shown in blue for curvature $\kappa(s,t)$. 
	(a) Two cases of reaching under no-obstacle setting with both rod and curvature profiles: Bend formation: The rod is initialized as straight and it forms a bend towards the target. Black arrows indicate the following: 1. the rod bends at the base and points towards the target; 2. the bend is formed; 3. the bend shifts from the mid-section to the tip; 4. the bend dissipates after the rod reaches the target. Bend propagation: The rod has an initial bend which propagates toward the target. Black arrows indicate the following: 1. the base curvature corrects the orientation of the rod toward the target; 2. the bend propagates. 
	(b) Two cases of reaching under obstacle setting:  The rod performs a simple reaching movement with one obstacle present.  Two obstacles create a scenario of mimicking the octopus reaching through a hole to get the target on the other side (compare with Fig.~\ref{fig:exp}(c)).}
	\label{fig:cases}
	\vspace{-10pt}
\end{figure*}


\section{Simulation results} \label{sec:numerics}

\begin{table}[!t]
	\centering
	\caption{Parameters for models and numeric simulation}
	\begin{tabular}{ccc}
		\hline
		\hline\noalign{\smallskip}
		Parameter & Description & Numerical value \\
		\hline\noalign{\smallskip}
		\multicolumn{3}{c}{{\bf Rod model}}\\
		$L_0$ & length of the undeformed rod [cm] & $20$ \\
		$\rodbase$ & rod base radius [cm] & $1$ \\
		$\rodtip$ & rod tip radius [cm] & $0.1$ \\
		$\rho$ & density [kg/${\text{m}}^3$] & $1042$ \\
		$\zeta$ & damping coefficient [kg/s]  & $0.01$\\
		$E$ & Young's modulus [kPa] & $10$ \\
		\hline\noalign{\smallskip}
		\multicolumn{3}{c}{{\bf Drag model}}\\
		$\rhow$ & water density [kg/${\text{m}}^3$] & $1022$ \\
		$c\mytan$ & tangential drag coefficient & $0.155$ \\
		$c\myper$ & normal drag coefficient & $5.065$ \\
		\hline 
	\end{tabular}
	\label{tab:num_para}
	\vspace*{-10pt}
\end{table}

In this section, we show numerical simulations of rod movements under our proposed sensory feedback control law. The rod dynamics are solved using the open-source software \textit{Elastica}~\cite{gazzola2018forward, zhang2019modeling}. In all simulations, the variable $\radius(s) = \rodtip s/L_0 + \rodbase(1 - s/L_0)$ gives the radius profile of a tapered rod, based on measurements in real octopuses, 
$A (s) =\pi(\radius(s))^2$ and $I(s)=\frac{A(s)^2}{4\pi}$ are the cross sectional area and second moment of area, respectively. For stability of numerical simulations, a smooth function is used to approximate the indicator function in \eqref{eq:ctrl_law_mu}.
Parameter values used in simulations are reported in Table~\ref{tab:num_para}.
%

\subsection{Reaching a static target} \label{sec:numerics_static_target}

\noindent
\textit{1) Bend formation:} The rod is initialized to be straight. A static target is presented right above the mid-section of the rod. As can be seen from Fig.~\ref{fig:cases}(a) on the left column, the rod creates a bend at the arm mid-section (which is also the closest to the target) oriented towards the target. Once the bend is created, the control propagates the bend towards the target while reorienting the arm at the base. The local increase of curvature at the base stops when the target is reached, at which point the equilibrium configuration is obtained. The part of the rod between the bend point and the tip remains passive throughout the whole movement.
This case is an example of the target within the reach of the arm (Proposition~\ref{prop:catching}(i)).

\smallskip
\noindent
\textit{2) Bend propagation:} The rod is initialized with a bent configuration which is commonly seen in octopus arms. A static target is presented in the direction the initial bend is pointing to. The bend, initially close to the base of the rod, is then propagated along the arm until the rod stabilizes in a configuration that points toward the target (see Fig.~\ref{fig:cases}(a), right column). 
This case is an example of the target outside the reach of the arm (Proposition~\ref{prop:catching}(ii)).


\smallskip
\noindent
\textit{3) Reaching in the presence of one obstacle:} The rod is initialized to be straight. A static target is above the rod with an obstacle between the arm and the target. At first, the rod forms a bend as described in the first case above. When it encounters the obstacle, the passive portion of the rod slides past the obstacle, eventually reaching the target (see Fig.~\ref{fig:cases}(b), left column).

\smallskip
\noindent
\textit{4) Squeezing through a hole:} In this case, two obstacles are used to create a hole, mimicking the experiment illustrated in Fig.~\ref{fig:exp}(c). The rod is initialized to be straight along the vertical direction. A static target is located on the other side of the obstacles. The same feedback control successfully drives the rod through the hole, squeezing and reaching to the target (see Fig.~\ref{fig:cases}(b), right column).

\subsection{Pursuing a moving target} \label{sec:simulation_moving_target}

\begin{figure}[!t]
	\centering
	\includegraphics[width=\columnwidth,trim={0pt 0pt 0pt 0pt}]{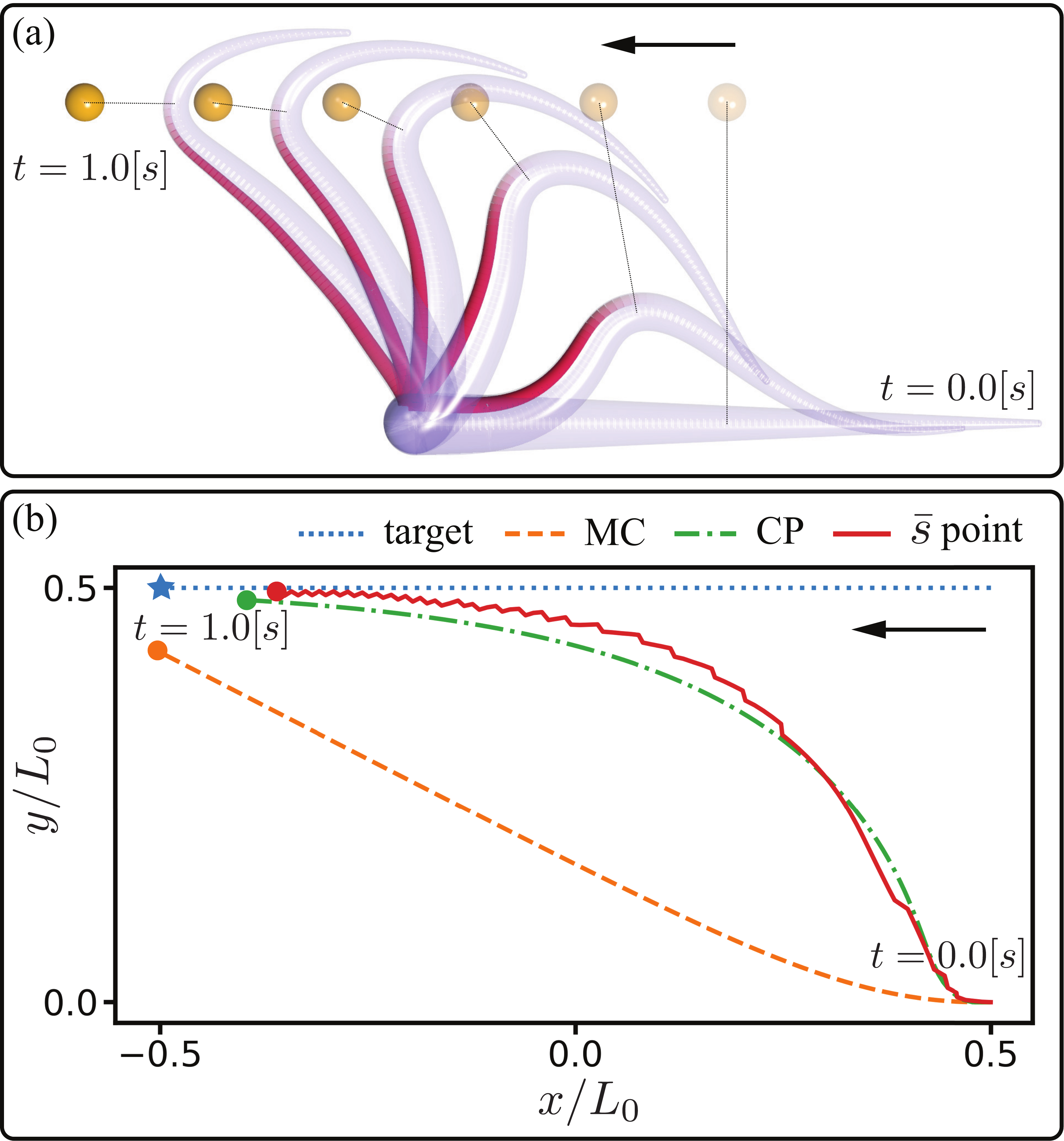}
	\caption{(a) The simulation result of the rod reaching a moving target: The target moves from the right to the left as indicated by the black arrow and is shown as a fade-in yellow sphere over six time instances (from $t = 0.0$ [s] to $t = 1.0$ [s]). (b) Comparison of trajectories: Three trajectories are compared for different pursuing behaviors. Red solid line represents the trajectory of the point of the rod ($\bar{s}$) closest to the target under the proposed sensory feedback control law~\eqref{eq:ctrl_law_rod}-\eqref{eq:ctrl_law_mu}. Orange dashed line denotes the trajectory of a pursuer under motion camouflage strategy~\eqref{eq:ctrl_law_MC}. Green dash-dotted line marks the trajectory of a pursuer under classical pursuit strategy~\eqref{eq:ctrl_law_CP}. The same target from (a) is shown here as a blue star with blue dotted trajectory. }
	\label{fig:comp-traj}
	\vspace{-10pt}
\end{figure}

The proposed sensory feedback control law~\eqref{eq:ctrl_law_rod}-\eqref{eq:ctrl_law_mu} is also capable of pursuing a moving target. A target is initiated above the mid-section of the rod and moved with constant speed ($0.2$ [m/s]) toward the left. Fig.~\ref{fig:comp-traj}(a) illustrates the rod pursuing the target. A bend is first created and is then propagated towards the tip, as the arm tries to follow the target.

We compare the trajectory of the $\bar{s}$ point (the closest point to the target) with other two reference trajectories. One is a pursuit trajectory using motion camouflage strategy (MC), while the other employs classical pursuit strategy (CP). The pursuer dynamics is given by~\eqref{eq:dynamics_unicycle} with constant speed which is the average speed of the $\bar{s}$ point from rod simulation. For both motion camouflage and classical pursuit cases, we take the control parameter $\chi=25$ in control laws~\eqref{eq:ctrl_law_MC} and~\eqref{eq:ctrl_law_CP} (see Appendix~\ref{appdx:pursuit}).

As depicted in Fig.~\ref{fig:comp-traj}(b), the trajectory of the $\bar{s}$ point qualitatively matches the trajectory of classical pursuit strategy rather than the motion camouflage stragety. This comparison justifies our hypothesis (see Remark~\ref{remark:dynamic_CP}) that the closest point along the rod follows a classical pursuit trajectory to chase the target.

\section{Comparison with bend propagation experiments} \label{sec:experiment}

\begin{figure*}[!t]
	\centering
	\includegraphics[width=\textwidth, trim={0pt 0pt 0pt 0pt}]{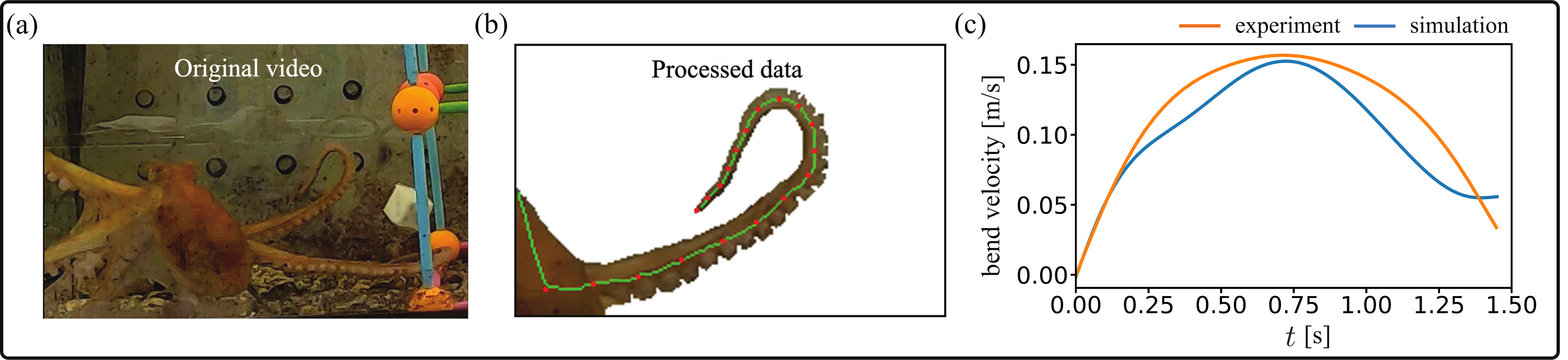}
	\caption{(a) An image frame from the original video of an octopus doing bend propagation (also see Fig.~\ref{fig:exp}(b)). (b) The original image is processed by background subtraction, binarization, skeletonization to get an isolated arm with estimated centerline (in green). Then we manually mark the data points (in red) along the estimated centerline  according to certain suckers' positions. The marked data points are used for smooth arm reconstruction. (c) A quantitative comparison of bend velocity profile between the reconstructed octopus arm from experiment and the arm from simulation. Both the velocity profiles are bell shaped.}
	\label{fig:exp-compare}
	\vspace{-15pt}
\end{figure*}

In this section, we provide a comparison between results from the proposed sensory feedback control and experimental data of the bend propagation movement depicted in Fig.~\ref{fig:exp}(b). We used a two-camera system for video recording. The experimental setup is similar to that of \cite{yekutieli2007analyzing, kim2022physics, wang2022control}.


\subsection{Arm reconstruction}


\smallskip
\noindent
\textit{1) Data processing:} The image analysis software Fiji \cite{schindelin2012fiji} was used for processing videos of arm reaching behaviors. Obtaining the centerline of the arm from video frames requires the following steps: (i) background subtraction, (ii) conversion to black-and-white images (binarization), and (iii) skeletonization. This process of estimating the centerline of the arm through skeletonization, is similar to previous octopus arm tracking examples~\cite{kazakidi2015vision, yekutieli2007analyzing}. Finally, data points along the arm are manually marked for each video frame, based on the locations of reference suckers, after overlaying the obtained centerline on top of the arm in the background-subtracted frames (see Fig.~\ref{fig:exp-compare}(b)). Each data point represents the location of the same sucker across all video frames. 

\smallskip
\noindent
\textit{2) Data smoothing:} In \cite{kim2022physics}, a smoothing algorithm was proposed to estimate all six modes of deformation from marked data points on a slender body. The same algorithm is used here, restricted to the planar case. 


\subsection{Bend velocity profile}
In bend propagation movement, the peak of the curvature profile provides the location of the bend point. We extract the velocity of the bend point from the smooth reconstructed arm (over forty-four frames). 
We also calculate the bend velocity profile from the bend propagation simulation presented in Sec.~\ref{sec:numerics_static_target} (Fig.~\ref{fig:cases}(a), right column). Noisy bend velocity profiles are smoothed using a low-pass filter (cut-off frequency of 1 Hz). Both of these profiles resemble a bell curve as shown in Fig.~\ref{fig:exp-compare}(c). Notably, such bell-shaped velocity profiles are prominently highlighted in octopus literature \cite{gutfreund1998patterns, sumbre2001control}, and are naturally recovered here as outcome of our control law.


\section{Conclusion and Future Work} \label{sec:conclusion}

In this paper, a novel sensory feedback control law is described for octopus arm movements.  The sensor model is motivated by behavioral experiments involving suckers.  The control law is inspired in equal part by experimental observations of bend propagation and pursuit strategies studied in literature.  For the proposed control law, analytical results are obtained including characterization of the equilibrium and its stability analysis. Several numerical simulation results show life-like motions of the soft arm. 

A salient contribution is the numerical comparison against the experimental data of bend propagation in a freely moving octopus arm. The control law is shown to recover the characteristic bell shape of the bend velocity profile.  The comparison helps provide the first such explanation of bend propagation using a feedback control law.

Future work will consider more realistic sensory models, including partially observed chemosensory and proprioceptory signals, and extend our analysis to 3D scenarios.

\bibliographystyle{IEEEtran}
\bibliography{reference}

\appendices
\renewcommand{\thelemma}{A-\arabic{section}.\arabic{lemma}}
\renewcommand{\thetheorem}{A-\arabic{section}.\arabic{theorem}}
\renewcommand{\theequation}{A-\arabic{equation}}
\renewcommand{\thedefinition}{A-\arabic{definition}}
\setcounter{lemma}{0}
\setcounter{theorem}{0}
\setcounter{equation}{0}



\section{Drag model}\label{appdx:drag}
The drag model is closely based upon~\cite{yekutieli2005dynamic, wang2022control}.  
We write the drag forces as
\begin{equation}
	\mathbf{f}^{\text{drag}} = -\frac{1}{2}\varrho_{\text{water}} \mathsf{Q} \begin{bmatrix}
		A\mytan c\mytan v_1|v_1| \\ 
		A\myper c\myper v_2|v_2|
	\end{bmatrix}
\end{equation} 
where $\varrho_{\text{water}}$ is the density of water, $A\mytan(s)=2\pi \radius(s)$ is 
the surface area of a unit length segment, and $A\myper=2\radius(s)$ is 
the projected area of the unit length segment in the plane perpendicular 
to the normal direction. Here $\radius(s)$ denotes the radius of the circular cross section of the rod.  
The coefficients $c\mytan$ and 
$c\myper$ denote the tangential and perpendicular drag 
coefficients.  Typically, $c\myper$ is much larger 
than $c\mytan$. Finally, $v_1$ and $v_2$ are the components of the velocity $\mathbf{r}_t$ in the material frame, i.e., $\mathbf{r}_t = v_1\mathbf{a} + v_2\mathbf{b}$.

\section{Proofs}\label{appdx:proof}

\subsection{Proof of Proposition~\ref{prop:catching}}\label{appdx:catching}
At the outset, define $\Gamma(s) :=\dist_s=-\cos\alpha$ and denote $\dist_0=\dist(0)$. 

(i) For this case, we necessarily have $\dist_0\leq L_0$. Here $\bar{s}\leq L_0$ is given. The proof is completed in the following two steps:

\noindent \textit{Step 1:}
Define $\dist_1\in(0,\dist_0)$ such that $\tilde{\mu}=\frac{1}{\dist_1}+c$ for some $c>0$. Then,
\begin{equation*}
\Gamma_s=-\sin^2\alpha\left(\tilde{\mu}-\frac{1}{\dist}\right) \leq -c(1 - \Gamma^2),\quad \forall \dist\geq\dist_1
\end{equation*}
Note that $\dist_s=-\cos\alpha\geq-1$, which implies
\begin{equation*}
\begin{aligned}
\dist(s)\geq\dist_1\,\ \forall s\leq \dist_0-\dist_1=:s_1
\end{aligned}
\end{equation*}
Therefore, it is guaranteed that $\Gamma_s\leq0,\ \forall s\leq s_1$. By separation of variables and some calculations, we may derive
\begin{equation*}
\Gamma(s) \leq \tanh\left(\tanh^{-1}\Gamma_0 - cs\right), ~~ \forall s \leq s_1
\end{equation*}
where we denote $\Gamma_0=\Gamma(0)$. We can therefore conclude that $\Gamma(s_1)\leq\tanh\left(\tanh^{-1}\Gamma_0 - cs_1\right)$. 

Note that for some $\epsilon_1 >0$ sufficiently small, $\tanh(z)\leq-1+\epsilon_1\Leftrightarrow z\leq\frac{1}{2}\ln\left(\frac{\epsilon_1}{2-\epsilon_1}\right)=:z_0$. Thus, if we take $c$ to be sufficiently large such that
\begin{equation*}
c \geq \frac{\tanh^{-1}\Gamma_0 - z_0}{s_1} =: c_1
\end{equation*}
then we are guaranteed to achieve $\Gamma(s_1)\leq-1+\epsilon_1$ given any small $\epsilon_1>0$ by using large enough $\tilde{\mu}=\frac{1}{\dist_1}+c$.

\smallskip
\noindent \textit{Step 2:}
Note that
\begin{equation*}
	\begin{aligned}
		\Gamma_s &= -\sin^2\alpha\left(\tilde{\mu}-\frac{1}{\dist}\right) 
		= -\sin^2\alpha\left(\frac{1}{\dist_1}+c-\frac{1}{\dist}\right)\\
		&\leq 0, ~~ 
		\forall \ \dist \geq \frac{\dist_1}{c\dist_1 + 1} =: \dist_2
	\end{aligned}
\end{equation*}
Similar to Step 1, we have $\dist(s)\geq\dist_2\,\ \forall s\leq \dist_0-\dist_2=:s_2$. For any $\epsilon>0$, choose $\dist_2\leq\epsilon$, i.e.,
\begin{equation*}
	c \geq \frac{1}{\epsilon} - \frac{1}{\dist_1} =: c_2
\end{equation*}
Then, by taking $c\geq \max\{c_1, c_2\}$, we have $\dist_s=\Gamma(s)\leq \Gamma(s_1) \leq -1+\epsilon_1,\ \forall s_1\leq s\leq s_2$. Then, for any $\epsilon>0$, we have
\begin{equation*}
\footnotesize
	\begin{aligned}
		\dist(s_2) &= \dist_0 + \int_0^{s_1}\dist_s\ud s + \int_{s_1}^{s_2}\dist_s\ud s \\
		&\leq \dist_0 + \int_0^{s_1} \tanh\left(\tanh^{-1}\Gamma_0 - cs\right) \ud s + (-1 + \epsilon_1)(s_2 - s_1) \\
		&\leq \underbrace{\dist_0 + \frac{\ln\left(\frac{\cosh(b)}{\cosh(b-cs_1)}\right)}{c}}_{h(c)} + (-1+\epsilon_1)\frac{c\dist_1^2}{c\dist_1+1} \leq \epsilon
	\end{aligned}
\normalsize	
\end{equation*}
where $b = \tanh^{-1} \Gamma_0$. One can derive that $h'(c)<0$ for $c\geq c_1$ and $h(c)\rightarrow \dist_1$ for $c\rightarrow \infty$.
Then, $\exists c_3 >0$ s.t. $\dist(s_2)\leq\epsilon$ for $c\geq c_3$ and small enough $\epsilon_1$.

Note that for $s\in[s_2,\bar{s}]$, we have $\tilde{\mu}-\frac{1}{\dist(s)}\leq 0$, $\alpha(s)\in(0,\frac{\pi}{2}]$ and thus, $\alpha_s>0$. Moreover, $\dist_s=-\cos\alpha\leq 0$ for $s\in[s_2,\bar{s}]$. Hence, $\dist(\bar{s})\leq\dist(s_2)\leq \epsilon$ by having $c\geq \max\{c_1,c_2,c_3\}$.

In conclusion, for all $\epsilon>0$, choose $\dist_1\in(0,\dist_0)$, $\exists \epsilon_1,c$ s.t. $\epsilon_1>0$ and $c\geq \max\{c_1,c_2,c_3\}$, i.e., $\exists\tilde{\mu}=\frac{1}{\dist_1}+c>0$ large enough so that $\dist(\bar{s})\leq\epsilon$ for given $\bar{s}\leq L_0$.

\smallskip
(ii) For this case, we have $\dist_0>L_0$ and $\bar{s}=L_0$. The proof is immediate by Step 1 in case (i) by choosing $\dist_1=\dist_0-L_0$, i.e. $s_1 = \bar{s} = L_0$.

\subsection{Proof of Proposition~\ref{prop:equilibrium}}\label{appdx:equilibrium}
In \cite[Sec. III-E]{chang2021controlling} it has been shown that if the internal muscle forces and couples are expressible as gradients of an  energy function (called \textit{muscle stored energy function}), then the system maintains its Hamiltonian structure (with damping) and (local) convergence to an equilibrium can be readily shown. In the present case, we see that the internal elastic couple is gradient of a quadratic elastic stored energy function $W^{\text{e}} = \tfrac{1}{2} EI \kappa^2$. Note that for any time $t$, given some curvature profile $\kappa$, the bearing angle $\alpha$ can be uniquely determined, i.e. we may express $\alpha = \alpha (\kappa)$. Then define $W^{\text{m}} = \int \mu \sin (\alpha (\kappa)) \, \dif \kappa$. Then it is clear that $u$ in \eqref{eq:ctrl_law_rod} is gradient of the function $W^{\text{m}}$. This completes the proof.

\section{Pursuit strategies for a unicycle}\label{appdx:pursuit}
Consider a point particle (pursuer) on a plane pursuing an evading target. The dynamics of a pursuer are described by the following unicycle system (states are the position $(\unicycle{x} (t), \unicycle{y}(t))$ and orientation $\thetaUnicycle (t)$):
\begin{equation}
\begin{aligned}
\dot{\unicycle{x}} = \unicycle{v} \cos \thetaUnicycle, ~~\dot{\unicycle{y}} = \unicycle{v} \sin \thetaUnicycle, ~~
\dot{\thetaUnicycle} = \unicycle{u}
\end{aligned}
\label{eq:dynamics_unicycle}
\end{equation}
Here, the dot notation is used for time derivatives. Assume the pursuer moves at a constant speed $\unicycle{v}$ and the only control is the steering rate $\unicycle{u}$. The moving target's dynamics can be represented in a similar way. We assume the target is also moving at a constant speed $\unicycle{v}^\target$.

Let $\sigma (t)$ be the distance between the pursuer and the target, $\phi (t)$ is the bearing angle to the target, and $\psi (t)$ is the bearing angle of the pursuer with respect to the target. Then the time evolution of $(\sigma, \phi)$ can be written as \cite{halder2016steering}
\begin{align}
\begin{split}
\dot{\sigma} &= - \unicycle{v} \cos \phi - \unicycle{v}^\target \cos \psi \\
\dot{\phi}     &= - \unicycle{u} + \frac{1}{\sigma} \left( \unicycle{v} \sin \phi + \unicycle{v}^\target \sin \psi \right) 
\end{split}
\label{eq:bearing_dynamics}
\end{align}

The motion camouflage control law \cite{justh2006steering} is the steering control given by
\begin{equation}
\unicycle{u} = \chi \left(\sin\phi + \frac{\unicycle{v}^\target}{\unicycle{v}}\sin\psi  \right)
\label{eq:ctrl_law_MC}
\end{equation}
where $\chi>0$ is some large enough given constant.

The classical pursuit control law \cite{galloway2013symmetry, halder2016steering} is the steering control given by
\begin{equation}
\unicycle{u} =  \chi \sin\phi  + \frac{1}{\sigma}\left(\sin \phi + \frac{\unicycle{v}^\target}{\unicycle{v}}\sin \psi \right)
\label{eq:ctrl_law_CP}
\end{equation}
where $\chi>0$ is some large enough given constant.



\end{document}